# Efficient Support of Big Data Storage Systems on the Cloud

Akshay MS, Suhas Mohan, Vincent Kuri, Dinkar Sitaram, H. L. Phalachandra
PES Institute of Technology, CSE Dept., Center for Cloud Computing and Big Data
dinkars@pes.edu

**Abstract**

Due to its advantages over traditional data centers, there has been a rapid growth in the usage of cloud infrastructures. These include public clouds (e.g., Amazon EC2), or private clouds, such as clouds deployed using Open-stack. A common factor in many of the well-known infrastructures, for example Openstack and Cloudstack, is that networked storage is used for storage of persistent data. However, traditional Big Data systems, including Hadoop, store data in commodity local storage for reasons of high performance and low cost. We present an architecture for supporting Hadoop on Openstack using local storage. Subsequently, we use benchmarks on Openstack and Amazon to show that for supporting Hadoop, local storage has better performance and lower cost. We conclude that cloud systems should support local storage for persistent data (in addition to networked storage) so as to provide efficient support for Hadoop and other Big Data systems

*Categories and Subject Descriptors*   D.4.2 [**Operating Systems**] Storage management – secondary storage, and D.4.7 [**Operating Systems**] Organization and design – distributed systems.

## 1. Introduction

In the recent past, there has been a widespread growth in the use of cloud infrastructures. The major reason for this growth is that in general, it is more efficient and less expensive to host applications on the cloud. Since these considerations apply also to Big Data systems such as Hadoop, it is important to support them efficiently on the cloud. For example, a major factor driving the adoption of cloud technology is resource sharing [Cr2009]. Since the demands of applications are typically bursty, it is possible to share the same server resources between multiple applications, leading to lower costs. Other advantages include the ability to scale server resources rapidly, and to have large spare capacity [Ar2010]. These factors indicate that it is important to support Big Data systems efficiently on cloud infrastructures.

Big Data systems by definition operate on large datasets. Therefore, when designing support for such systems on a cloud system, it is important to consider the storage architecture of the cloud system. Hadoop, Google File System, and other Big Data systems use local storage for reasons of high performance and low cost [Bo2008, GH2003]. Cloud systems, however, typically use networked storage for persistent data. For example, Openstack supports only networked storage (e.g., ISCSI) for persistent data [Ci2013]. Cloudstack, which is another widely used cloud system, supports networked storage together with local storage [Ci2012]. However, local storage is recommended only for non-persistent storage [Lo2012].

One of the motivations for using networked storage in cloud systems is to provide for availability of the data in the face of failure. This consideration does not apply to many Big Data systems including Hadoop since they replicate the data. Thus if one copy of the data is lost (either due to server or disk failure), other copies are still available.

The rest of the paper is organized as follows. Section 2 reviews related work in this area. Our local storage-based solution is presented in Section 3, followed by benchmark results in Section 4. Section 5 contains our conclusions.

## 2. Related Work

There are a number of ways in which Big Data systems can be supported in a cloud. Networked storage is used to support persistent data in Openstack and other cloud systems [Ci2013], and can be leveraged to store Hadoop data. We use benchmarks to show our solution has higher performance. Another alternative is to use a high performance storage technology, for example Amazon EC2 Elastic Block Storage [EB2013]. While this solution can achieve high performance, we provide data to show that it is not as low cost as our proposal.

There are many studies of Hadoop performance [Sh2010, Su2010, Xi2010]; however, these concentrate on the performance of Hadoop running on bare hardware. The work that comes closest to ours is [Sh2010a] which uses Hadoop running on Eucalyptus and Amazon to identify virtualization bottlenecks in the Eucalyptus and Amazon cloud systems. However, the focus of the paper is on improvement of virtualization technology and not upon an efficient design to support Hadoop on cloud systems.

## 3. Our Solution

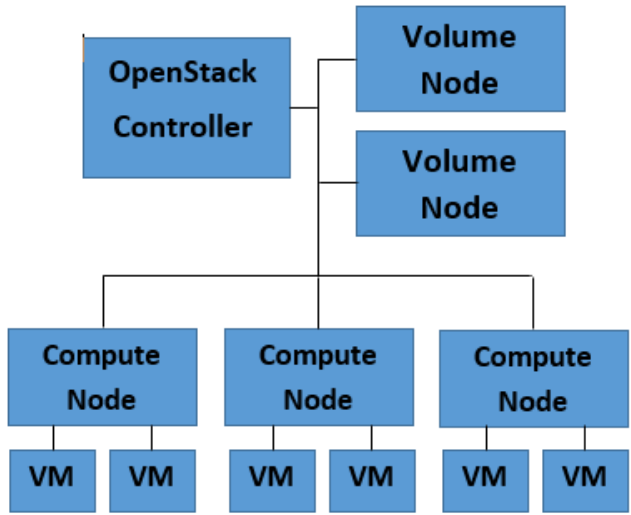

Figure 1: Openstack System with Hadoop VMs

Our objective is to provide a high-performance Hadoop system running on a cloud infrastructure. We first describe the design considerations that arise, followed by the description of the actual solution.

### 3.1 Design Considerations

The design approach of Big Data systems is moving the program to the data and not the data to the program. This ensures that data processing takes most of the time and not network transfer of data to the machines where the jobs are being deployed [Bo2008, GH2003]. A network attached storage system will not be able to exploit the full capabilities of Hadoop unless the storage system is connected via a very high speed network – 10G or Fiber Channel which makes the entire setup expensive and requires considerable expertise to set up and maintain the cluster. We note that even with high-performance networking, disk performance cannot be higher than local storage since even iSCSI, data eventually comes off disk that is directly attached to another machine

Therefore, in our solution, we propose that running Hadoop on local storage is ideal from a cost and performance perspective. The major constraint with current cloud systems such as Openstack is that the data stored in local storage is not persistent. There are two methods of getting around this – our current method relies on having long-running VMs so that the local storage is available for a long time. The second method is to extend cloud with persistent local storage.

Having Hadoop data on local storage has the additional advantage that it inter-operates better with Hadoop load-balancing algorithms. When scheduling tasks, Hadoop load-balancing algorithms try to factor in *data locality*, i.e., information about the nodes on which data resides. These load balancing algorithms will not work well with network attached disks. Additionally, VM migration for balancing load, which is commonly used in cloud systems, is probably not useful for balancing the load since it does not take these factors including data locality into account. Therefore, migration of Hadoop VMs should be disabled.

A final consideration arises from the objective of making sure that Hadoop's replication facility is not inadvertantly defeated by running Hadoop on a cloud infrastructure. In a cloud, it is possible that all 3 VMs containing the replicas of a file would be scheduled on the same physical machine. To prevent this, we use Hadoop's *rack awareness* property, All VMs running on the same physical machine are designated (to Hadoop) as being in the same (virtual) rack. Hadoop would then ensure that there are at least two different replicas across racks; i.e., that there are two replicas in different physical machines. Since rack awareness is a common feature of Big Data systems, this method can be used for other Big Data systems as well.

### 3.2 Details of Our Solution

Figure 1 contains a high-level overview of our solution. We have a single controller node running core OpenStack services such as Keystone, Glance, Cinder and Quantum. Cinder is the volume management service and volumes created using Cinder reside on the controller and are attached to the virtual machines over iSCSI. We have several compute nodes running nova-compute service that can spawn virtual machines. Each physical node has Intel Xeon E3-1220 v2 @ 3.10GHz, 8MB Cache with 16 GB RAM and 1 TB Hard disk. All the nodes are connected to two different networks - 1Gbps each. One network is used for OpenStack services to communicate with each other and the other is used to connected to a public network. A number of long-running Hadoop VMs are spawned on OpenStack. These VMs are similar in behaviour to a Hadoop cluster, with each VM being similar to a Hadoop node. Spawning more VMs than needed is not a performance overhead since the VMs consume very little resources when they are not active.

OpenStack instances can have three types of storage - a root disk, an ephemeral disk that is non-persistent and persistent storage attached over the network through OpenStack's volume service. The root disk of a virtual machine resides on the host machine and it is not attached over the network. This implies that the root disk of the virtual machine does not depend on the network latency or bandwidth. Our solution to run Hadoop involves using the root disk for HDFS as shown in Figure 2. This avoids transfer of data over the network while running Hadoop jobs and is cost effective. For the Amazon comparison in Section 4.2,

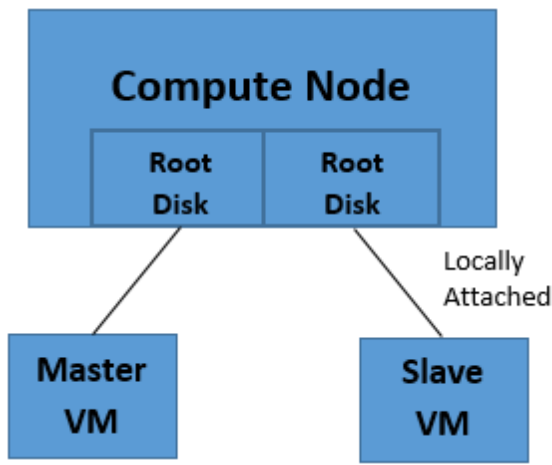

Figure 2: Hadoop VMs with Local Storage

$$Throughput(N) = \frac{\sum_{i=0}^{N} filesize_i}{\sum_{i=0}^{N} time_i}$$

$$Average\ IO\ Rate = \frac{\sum_{i=0}^{N} rate_i}{N} = \frac{\sum_{i=0}^{N} \frac{filesize_i}{time_i}}{N}$$

Figure 3: Definitions of Throughput and Average I/O Rate

we do a similar setup by using the the instance storage of the instance. Instance storage on EC2 is non-persistent but we have found it to be faster and cheaper than standard EBS volumes.

Since the root disk is not persistent, i.e. data stored on the root disk is lost after the VM is terminated, we need to periodically snapshot the data in the root. This can be performed as an asynchronous background task. The overhead in snapshots is generally lower than the overhead of accessing all I/O via the network. Most Big Data systems (e.g., Google search) write data once, but read it many times. Since only the writes have to be snapshotted, the overhead is lower. In practice, we also found that the storage does not disappear immediately if VM crashes, so that if the VM can be re-booted quickly, the storage will not be lost.

### 3.3 Disk Partitioning Solution

Both the root disk and the ephemeral disk in Openstack are implemented as files on the local storage. An alternative would be to partition the local storage disks, and attach one or more of the partitions to the Hadoop VMs. This would give higher performance than our current solution, and is under implementation. The disadvantage of doing so is that a static partition of the disk would be dedicated to the Hadoop VM, whereas with the current implementation of root and ephemeral disks, it is possible for the amount of storage allocated to these disks to shrink and grow. Nevertheless, we intend to experiment with this alternative solution, as we believe that the gains in performance may outweigh the loss of flexibility for some applications.

Implementation of the above solution is also not difficult in the current Openstack architecture. We assume that this storage is a new type of storage called *local-persistent*. It would be necessary to implement a new Openstack component that would keep track of the local-persistent partitions. Currently, Openstack contains a configuration flag *libvirt_images_volume_group* that specifies, on each compute node, the volume group that contains ephemeral disks. We plan to add a similar flag *libvirt_local-persistent_volume_group* that contains local persistent volumes. Access to these volumes would be via the usual Openstack access control mechanisms. Long-running Hadoop VMs could be started only on compute nodes that contain local-persistent storage using the Openstack filter scheduler. This scheduler allows the administrator to filter the list of nodes on which a new VM is launched. The VM initialization sequence has also got to be modified to avoid formatting the local-persistent disks attached to the instance.

### 4. Benchmarks and Measurements

TestDFSIO is a standard benchmark used for testing the I/O performance of a Hadoop system [Mi2011]. In the following, we describe TestDFSIO, followed by a comparison of results of running this benchmark using our solution, standard Openstack, and Amazon. The comparison includes both performance and price comparisons of our solution with standard Openstack and Amazon.

### 4.1 TestDFSIO

The operation of TestDFSIO is a distributed I/O benchmark that works as follows. When TestDFSIO is invoked on a Hadoop cluster, it invokes the MapReduce infrastructure to create a number of parallel tasks on each node (shown diagrammatically in Figure 6). The benchmark, therefore, simulates the operation of a real Hadoop task. Each parallel task does I/O to a separate file at the maximum possible rate. The I/Os can be writes, reads or a mixture of reads and writes. It is conventional to run TestDFSIO and measure the write performance first, so as to create the files for subsequent measurements of read performance [Mi2011]. This test writes into or reads from a specified number of files. File size is specified as a parameter to the test. Each file is accessed in a separate map task [Te2013].

The reducer collects the following statistics:
- Number of tasks completed
- Number of bytes written/read
- Execution time
- I/O rate
- I/O rate squared

The following statics are obtained after the job is completed:
- Read or write test
- Date and time the test finished
- Number of files
- Total number of bytes processed
- Throughput in MB/sec (total number of bytes / sum of processing times)
- Average I/O rate in MB/sec per file
- Standard deviation of I/O rate

TestDFSIO generates two important metrics. The *Throughput* is the total I/O by the cluster per unit time per node. For a TestDFSIO job using $N$ map tasks, and where the index $1 <= i <= N$ denotes the individual map tasks, the throughput is defined by the equation in Figure 2 [Mi2011]. The *Average IO Rate* measures the average I/O rate per node and is given by the equation in Figure 3. For $N$ identical nodes, the two values should be almost identi-

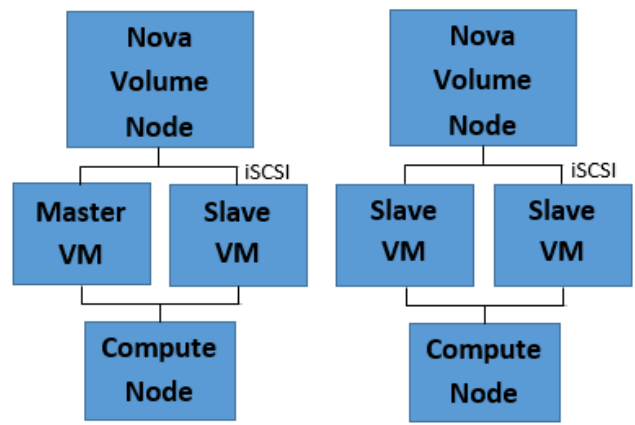

Figure 4: Standard Deployment of Hadoop on Openstack

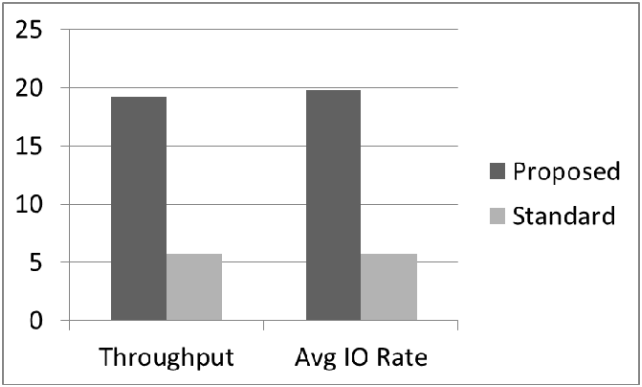

Figure 5: Comparison of Write Performance cal.

### 4.2 Performance Comparison

We first compare the performance of our solution against the standard method of deploying Hadoop on Openstack. TestDFSIO was run on a 5 node Hadoop cluster with a map capacity of 25. Each virtual machine we have used contains 4VCPUs, 8GB of RAM and 32 GB root disk and 20GB ephemeral storage. The 5 VMs are run on a 5 node OpeStack cluster i.e., each physical node hosts 1 VM. A total of 10 files, each of 1000MB were used to perform the benchmark. The configuration used in measuring our solution is shown in Figure 1. Figure 4 shows the standard method of deploying Hadoop on Openstack.

Figure 5 compares the write performance of our proposed solution using locally attached disks against the standard method of using iSCSI disks for storing persistent data. The Y-axis of the figure is in units of MB/s. It can be seen that there is a substantial difference in write performance. As expected, the Throughput and Average IO Rate figures are very close. For read performance, the Average IO Rate for the Proposed solution and the standard solution are 230 MB/s and 176 MB/s, respectively.

### 4.3 Cost Comparison

The performance comparison in the previous section shows that the performance of our solution is superior to the performance achievable with the standard method of deploying Hadoop on Openstack. It is possible to replace the iSCSI interconnect with a higher performance interconnect. In this section, we show that our solution is likely to be more cost effective than solutions that use such interconnects.

The high performance cloud storage that we use for a cost comparison of our solution is Amazon Elastic Block Store. We compare the Amazon EBS solution with an implementation of our solution on Amazon using Amazon EC2 root disks for storage of local data. While the exact implementation of Amazon EBS is not known, it is believed to be a cluster disk implementation [Bl2010]. As of this writing, *Standard* EBS disks are capable of supporting a steady I/O rate of upto 100 IOPS (I/O operations per second), with bursts of upto twice or thrice that rate. Additionally, Amazon also provides *Provisioned* EBS disks, which can support burst rates of up to 30,000 IOPS.

Detailed cost comparisons require taking many factors into account, for example the cost of hardware, and operational and maintenance costs. To provide an objective basis for such comparisons, we assume that the prices charged by Amazon for their services are indicative of the underlying costs of providing these services. For doing the cost comparison, we compare the costs of deploying two configurations on Amazon. The first configuration is similar to our solution, while the second solution leverages high-performance EBS disks for storing Hadoop data.

The details of the configurations are as follows. The nodes in the Hadoop cluster were first generation large instance (m1.large) with a 100GB EBS Standard volume attached. In our experiments, we have set up a 5 node Hadoop cluster on Amazon EC2 consisting of 1 master node and 4 slave nodes. The root disk is also a standard volume without provisioned IOPS. m1.large machines are known to give moderate IO performance [Am2013a]. The instance comes with 850 GB of ephemeral storage, which is storage that is locally attached to our machine. The first configuration used the ephemeral disk for HDFS. This is similar to our solution, since the ephemeral disk is local storage. The second configuration used standard EBS volume for HDFS data. This corresponds to using high-performance networked cloud storage.

Table 1: Services and their respective prices for AWS in North Virginia Region [Am2013]

| Instance | Cost |
| --- | --- |
| EC2 – M1.Large | $0.24 per hour |
| EBS – Standard | $0.10 per 1 millions IOPs |
| EBS-IOPs | $0.10 per IOPS-Month |

Table 1 lists the costs of various Amazon services. While running TestDFSIO, on AWS we were able to find the exact number of I/O operations performed on the disk, using Amazon detailed monitoring. Our test run requires slightly over 1 million I/O operations if run continuously for an hour. If Hadoop was run on EBS volumes, we would be charged $0.10 in addition to the cost of running the instance for 1 hour i.e., $0.24. Since using our proposed solution eliminates the need for EBS volumes, we can avoid the cost of running the volume. Therefore, using our solution of running Hadoop on Ephemeral disks in Amazon EC2 proves to be 29% cheaper than running it on EBS

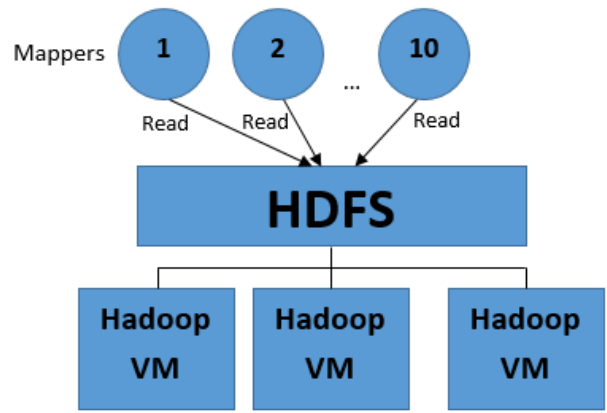

Figure 6: TestDFSIO Operation

volumes.

## 5. Conclusions

In this paper, we have shown that our solution for running Hadoop on a cloud infrastructure using local disks for persistent storage rather than networked storage has higher performance and is more cost-effective than the traditional alternatives. Based upon this, we argue that cloud infrastructures should support the use of persistent locally attached storage for efficient support of Big Data and other I/O intensive applications. The traditional argument for using networked storage for data availability does not apply to Big Data systems since they have replication and other availability methods already built in. Persistent local storage can co-exist with existing persistent network storage for other types of applications.

In our future work, we plan to extend our work to other applications, such as databases. We had also proposed to allow the attachment of disk partitions directly to VMs as proposed in Section 3.3. This extension of our solution would further improve the performance and efficiency.

## Acknowledgments

We would like to thank Abhishek B. S., Mahesh A., Rakesh Kumar, Sandeep Raju, Shruti Ranade, Vijesh M and Vivek P for their helpful discussions and comments.

## References

1. [Am2013] Amazon Elastic Cloud(EC2) Pricing http://aws.amazon.com/pricing/ec2/
2. [Am2013a] Amazon EC2 Instance Types http://aws.amazon.com/ec2/instance-types/
3. [Ar2010] Armbrust, Michael, Armando Fox, Rean Griffith, Anthony D. Joseph, Randy Katz, Andy Konwinski, Gunho Lee et al. "A view of cloud computing." Communications of the ACM 53, no. 4 (2010): 50-58.
4. [Bl2010] Bliekertz, Soren, "On Amazon EC2's Underlying Architecture" http://openfoo.org/blog/amazon_ec2_underlying_architecture.html
5. [Bo2008] Borthakur, Dhruba. "HDFS architecture guide." HADOOP APACHE PROJECT http://hadoop.apache.org/common/docs/current/hdfs design. pdf (2008).
6. [Ci2012] "Citrix XenServer Installation for CloudStack", Cloudstack Installation Guide, Chapter 8.2, http://incubator.apache.org/cloudstack/docs/en-US/Apache_CloudStack/4.0.0-incubating/html/Installation_Guide/citrix-xenserver-installation.html
7. [Ci2013] "Volumes." Openstack Compute Administration Manual, Chapter 11, http://docs.openstack.org/trunk/openstack-compute/admin/content/ch_volumes.html
8. [Cr2009] Creeger, Mache. "Cloud computing: An overview." ACM Queue 7, no. 5 (2009): 3-4.
9. [EB2013] "Amazon Elastic Block Store" http://aws.amazon.com/ebs/
10. [Gh2003] Ghemawat, Sanjay, Howard Gobioff, and Shun-Tak Leung. "The Google file system." In ACM SIGOPS Operating Systems Review, vol. 37, no. 5, pp. 29-43. ACM, 2003.
11. [Lo2012] "Local storage support for data volumes", Apache Cloudstack Project, https://cwiki.apache.org/CLOUDSTACK/local-storage-support-for-data-volumes.html
12. [Mi2011] Michael G. Noll "Benchmarking and Stress Testing an Hadoop Cluster with TeraSort, TestDFSIO, http://www.michael-noll.com/blog/2011/04/09/benchmarking-and-stress-testing-an-hadoop-cluster-with-terasort-testdfsio-nnbench-mrbench/#testdfsio
13. [Te2013] TestDFSIO.java http://svn.apache.org/repos/asf/hadoop/common/tags/release-1.2.0/src/test/org/apache/hadoop/fs/TestDFSIO.java
14. [Sh2010] Shafer, Jeffrey, Scott Rixner, and Alan L. Cox. "The Hadoop distributed filesystem: Balancing portability and performance." In Performance Analysis of Systems & Software (ISPASS), 2010 IEEE International Symposium on, pp. 122-133. IEEE, 2010.
15. [Sh2010a] Shafer, Jeffrey. "I/O virtualization bottlenecks in cloud computing today." In Proceedings of the 2nd conference on I/O virtualization, pp. 5-5. USENIX Association, 2010.
16. [Su2010] Sur, Sayantan, Hao Wang, Jian Huang, Xiangyong Ouyang, and Dhabaleswar K. Panda. "Can High-Performance Interconnects Benefit Hadoop Distributed File System." In Workshop on Micro Architectural Support for Virtualization, Data Center Computing, and Clouds (MASVDC). Held in Conjunction with MICRO. 2010.
17. [Xi2010] Xie, Jiong, Shu Yin, Xiaojun Ruan, Zhiyang Ding, Yun Tian, James Majors, Adam Manzanares, and Xiao Qin. "Improving mapreduce performance through data placement in heterogeneous hadoop clusters." In Parallel & Distributed Processing, Workshops and Phd Forum (IPDPSW), 2010 IEEE International Symposium on, pp. 1-9. IEEE, 2010.

.